\documentstyle[preprint,pre,aps]{revtex}
\tighten
\begin{document}
\draft

\title{Crossover of a nonequilibrium system
studied by coherent anomaly method}

\author{G\'eza \'Odor}

\address
{\it Research Institute for Materials Science, P.O.Box 49,
H-1525 Budapest, Hungary }

\date{\today}
\maketitle
\begin{abstract}

In the stochastic cellular automaton of rule 18 defined by S.
Wolfram [Rev.\ Mod.\ Phys.\ {\bf 55}, 601 (1983)]
the crossover of the critical behaviour induced by nonlocal
site exchange was investigated using the coherent
anomaly method. The continuous variation of the critical
$\beta$ exponent was confirmed.
\end{abstract}

\narrowtext

The addition of nonlocal exchange to models of short range
interaction has recently become a major topic of interest.
\cite{racz}
Long-range effective interactions can be defined by this
kind of dynamics. This is important in simulation
because time consuming interaction calculations can be
substituted by algorithms with simple nonlocal particle
exchange.

We have already shown \cite{od-sza} that a kind of long-range
mixing can be taken into account into analytical generalized
mean-field (GMF) \cite{gut,dick} calculations as well.
The stochastic cellular automaton (CA) of rule 18 \cite{boc}
with nonlocal site exchange was investigated and the
variation of the critical point and the order-parameter
exponent was determined by extrapolation and simple fitting.

The coherent anomaly method (CAM) \cite{suz} and its enhanced
version -- introduced very recently \cite{kol} -- have already
produced accurate critical exponents for equilibrium
statistical systems \cite{suz-min} and for dynamical systems
\cite{suz-kub,kat-suz,yah-suz,inui,odo}.
The motivation underlying the present work was to show that
the GMF + CAM method is able to describe the continuous
variation of the critical exponent in a dynamical system.
The results agree with former simulations and GMF
extrapolation data.

Cellular automata models are very useful tools for modelling
dynamical systems in many branches of science \cite{wolf}.
Even one-dimensional models possess very rich features
ranging from self-organized critical behaviour to chaotic
phenomena.
Since these models are non-equilibrium systems, they can
exhibit phase transitions even in one dimension.

The basic model to be investigated was the one-dimensional,
two-state, rule 18 CA \cite{wolf} with probabilistic
acceptance rate :
\[ s(t+1,j) = \left\{ \begin{array}{ll}

X & \mbox{if} \ \  s(t,j-1)=0, \ s(t,j)=0, \ s(t,j+1)=1, \\
X & \mbox{if} \ \  s(t,j-1)=1, \ s(t,j)=0, \ s(t,j+1)=0, \\
0 & \mbox{otherwise}
\end{array} \right. \]
where $X\in\{0,1\}$ is a two valued random variable such that
$$Pr(X=1)=p \ \ .$$
This synchronous update can be followed by a sequential
site exchange rule : $m\times$ the number of living cells
(`ones`) are selected randomly and swapped with other site
values (either zeros or ones) selected at random from all over
the lattice. This rule preserves the number of living cells.

Taking the steady-state concentration of ones as the
order-parameter $c$, the stochastic model without site
exchange exhibits a continuous phase transition to an empty
state if the acceptance probability $p$ is less than $p_c$.
This critical probability was found by steady-state simulation
to be $p_c = 0.8086(2)$ \cite{boc}. Time dependent simulation
\cite{gras} -- considered to give more accurate results
owing to the lack of finite size effects -- gave
$p_c = 0.8094(2)$ \cite{unp}.
Both simulations predicted that the universality of the
critical transition belonged to the directed percolation (DP)
class \cite{cardy}. The order-parameter exponent of this class
is $\beta\simeq 0.2769(2)$ \cite{dickj}.
Simple mean-field calculation results in $p_c^{MF} = 0.5$ and
$\beta_{MF} = 1$. The generalized mean field calculation, which
is based on the solution of steady-state equations for
$n$-block probabilities, gave a series of approximations
\cite{sza-od}.
This CA in the steady-state can be mapped to a simpler rule
(rule 6/16) with the new variables $01 \to 1$ and $00
\to 0$:
\begin{verbatim}

          t-1 :           0 0          0 1          1 0        1 1
          t :              0            1            1          0

\end{verbatim}
and the GMF equations can be set up by means of pair variables.
Figure \ref{fig1} shows the phase diagram of the model
calculated by  the GMF method and simulation. Pad\'e
extrapolation for the $n=6$ level GMF solution resulted
in $p_c = 0.7986$ and $\beta = 0.29$ \cite{sza-od}. CAM
extrapolation for the $n=7$ pair GMF solution gave an
estimate of $\beta = 0.2796(2)$ \cite{odo}.

If this stochastic CA model is supplemented with site exchange,
the correlations are partially destroyed and the transition
behaviour is shifted towards the simple classical mean-field
approximation results. We thus see a crossover of the
universality.
Steady state simulation for nearest neighbour site exchange
(local) and nonlocal exchange resulted in a similar continuous
crossover as a function of mixing strength, but on different
scales \cite{boc}. Though it is understandable that the
nonlocal exchange is much more effective than the local mixing,
it is not obvious for a nonlocal site exchange -- where a
particle can jump to any distance -- that the crossover is
still continuous.
To check this, simulations on different sized lattices and
GMF calculations were carried out. The mapping onto pair
variables is not possible in this case, since site exchange
destroys the ordered structure of the steady-state that -- it
cannot be built up from $00$ and $01$ pairs. GMF calculations
up to the $n=5$ level point correlation were performed; these
were followed by simple extrapolations \cite{od-sza}. The
variation of the $p_c(m)$ and the $\beta(m)$ was found to be
continuous in agreement with simulations. In what follows I
show that an independent extrapolation method (CAM) predicts
very similar results for the critical exponent, based on
finite size scaling .

Calculating critical exponents of second order phase
transitions is a challenging problem. Besides simulation
there are a number of analytical methods.
Theoretical tools for studying nonequilibrium statistical
physics processes are under development. The coherent
anomaly method introduced by Suzuki \cite{suz} is based
on a series of generalized mean-field approximations and
uses finite size scaling theory to estimate the quantities
$Q_n$ that depend on long-range correlations.

Quantities of the $n$-th level of approximation
in the vicinity of the critical point can be described
by the classical singular behaviour multiplied by an
anomaly factor ($a(n)$) :
\begin{equation}
Q_n \sim a(n) (p/p_c^n - 1)^{\omega_{cl}} \ ,
\end{equation}
where $p$ is the control parameter and $\omega_{cl}$
is the classical critical index.
The anomaly factor diverges in the
$n \to \infty$ (and $p_c^n \to p_c$) limit, but scales as :
\begin{equation}
a(n) \sim (p_c^n - p_c)^{\omega - \omega_{cl}}
\end{equation}
thereby permitting the estimation of the true critical
exponent $\omega$.

Many solved and unsolved equilibrium and nonequlibrium
critical systems have been studied by this method with success
\cite{suz-min}.
Recently a new parametrization was suggested instead of using
control parameter $p$,
\begin{equation}
\delta_n = (p_c/p_c^n)^{1/2} - (p_c^n/p_c)^{1/2} \ ,
\label{del}
\end{equation}
that has an invariance property: $p \leftrightarrow p^{-1}$.
Taking into account correction terms the anomaly scaling
now reads as :
\begin{equation}
a(n) = b \ \delta_n^{\beta - \beta_{cl}} +
       c \ \delta_n^{\beta - \beta_{cl} + 1} + ...
\label{corr}
\end{equation}

Using this `enhanced` CAM, very accurate critical indexes
were found for the 3 dimensional Ising model
\cite{kol}. The DP universality of the nonequilibrium
stochastic rule 18 CA was confirmed \cite{odo} with accurate
$\beta$ exponent estimation.

Here I used our $n = 1 ... 5$ point level GMF approximation
data with the effect of nonlocal site exchange included.
As was explained in \cite{od-sza} this mixing with $m << 1$
can be taken into account in the steady-state block
probabilities by considering the effects of single particle
"jump in" or "jump out" of a given $n$-block. For greater $m$
the mixing can be well described by iterating the one-particle
jump effect on the equations $k = m / m^,$ times such that
$m^,$ is small. This kind of calculation is in agreement with
the sequential simulation process. Our experience was that
the selection of $m^, \leq 0.001$ did not improve the results
therefore $m^, = 0.001$ was fixed through the calculations.

The anomaly factor was determined for each $m$ for
$n = 1 ... 5$.
\begin{equation}
a(n,m) = c(n,m) / (p/p_c(m) - 1)^{\beta_{MF}} \ ,
\label{nfit}
\end{equation}
where $c$ is the concentration, $\beta_{MF} = 1$.
The exponent $\beta$ was fitted out with non-linear regression
using formula (\ref{corr}) and the $p_c(m)$ values of
simulation. Alternatively one could use $p_c(m)$ estimates
determined from extrapolation of GMF data.
The stability of the fitting was checked by selecting
different subsets : $n = \{1,2,3,4,5\}$, $\{1,2,3,4\}$,
$\{1,2,5\}$ and $\{3,4,5\}$ of the GMF data. The
uncertanity was only a few percent for this model.

In Figure \ref{fig3} I have plotted the anomaly factors as
functions of $m$. The $n=1$ classical mean-field approximation
factor does not feel the effect of mixing and remains constant.
This is understandable since in the classical mean-field
approximation correlations are neglected and the mixing cannot
destroy them. Higher levels of approximation for $m \ge 1$
give the same anomaly factor value, which means according to
eq.(\ref{corr}) that the true critical behaviour is described
by $\beta = \beta_{MF} = 1$. As the mixing decreases from
$m=1$ the anomaly factors of $n > 1$ change continuously and
so does the exponent $\beta$. At $m=0$ the anomaly fators
$a(1,0), \ a(3,0), \ a(5,0)$ coincide with the anomaly
factors of the pair-correlation GMF results
$a(1), \ a(2), \ a(3)$ \cite{odo}.

Figure \ref{fig4} shows that the results obtained from this
CAM calculation agree very well with steady-state simulations
\cite{boc} and with earlier calculations \cite{od-sza}.
The crossover from $\beta_{MF} = 1$ begins at $m \simeq 1$.
The universality changes continuously toward the DP class
in the $m\to 0$ limit (characterised by $\beta_{DP}\simeq 0.28$ ).

In conclusion, the efficiency of the CAM method for describing
universality change in a nonequlibrium model is proven by this work.
The agreement with other results strengthens the belief that
nonlocal exchange causes continuous crossover in this model.
This method shows a possibility for studying models with
long-range effective interactions, generated by site exchange
dynamics.

This research was partially supported by the
Hungarian National Research Fund (OTKA) under grant numbers
T-4012 and F-7240.

\begin{figure}
\caption{Phase diagram of the stochastic rule 18 CA determined
from simulation (diamonds) and GMF pair-approximations at
levels $n=1,...,8$}
\label{fig1}
\end{figure}

\begin{figure}
\caption{Anomaly coefficients as functions of mixing
strength $m$ obtained from $n = \ 1(\triangle), \ 2(\times), \ 3(\Box), \ 4(+),
\ 5(\diamond) \ $ levels of GMF approximations.}
\label{fig3}
\end{figure}

\begin{figure}
\caption{Crossover induced by nonlocal site exchange $m$
from DP to mean-field universality. The phase diagram
was calculated by CAM ($+$) and steady-state simulation
($\times $). The DP value of $\beta_c \simeq 0.276$ is also shown.}
\label{fig4}
\end{figure}

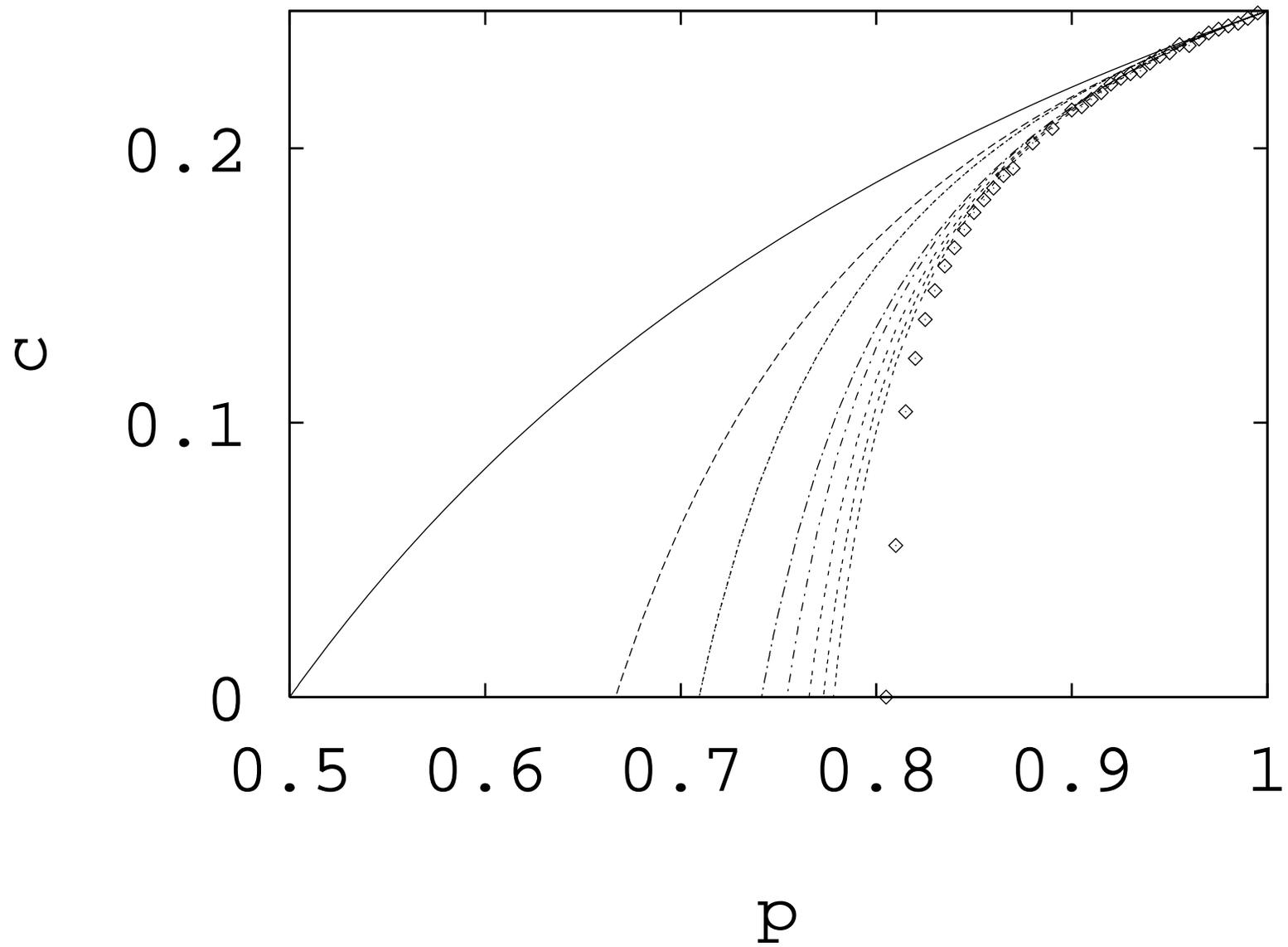
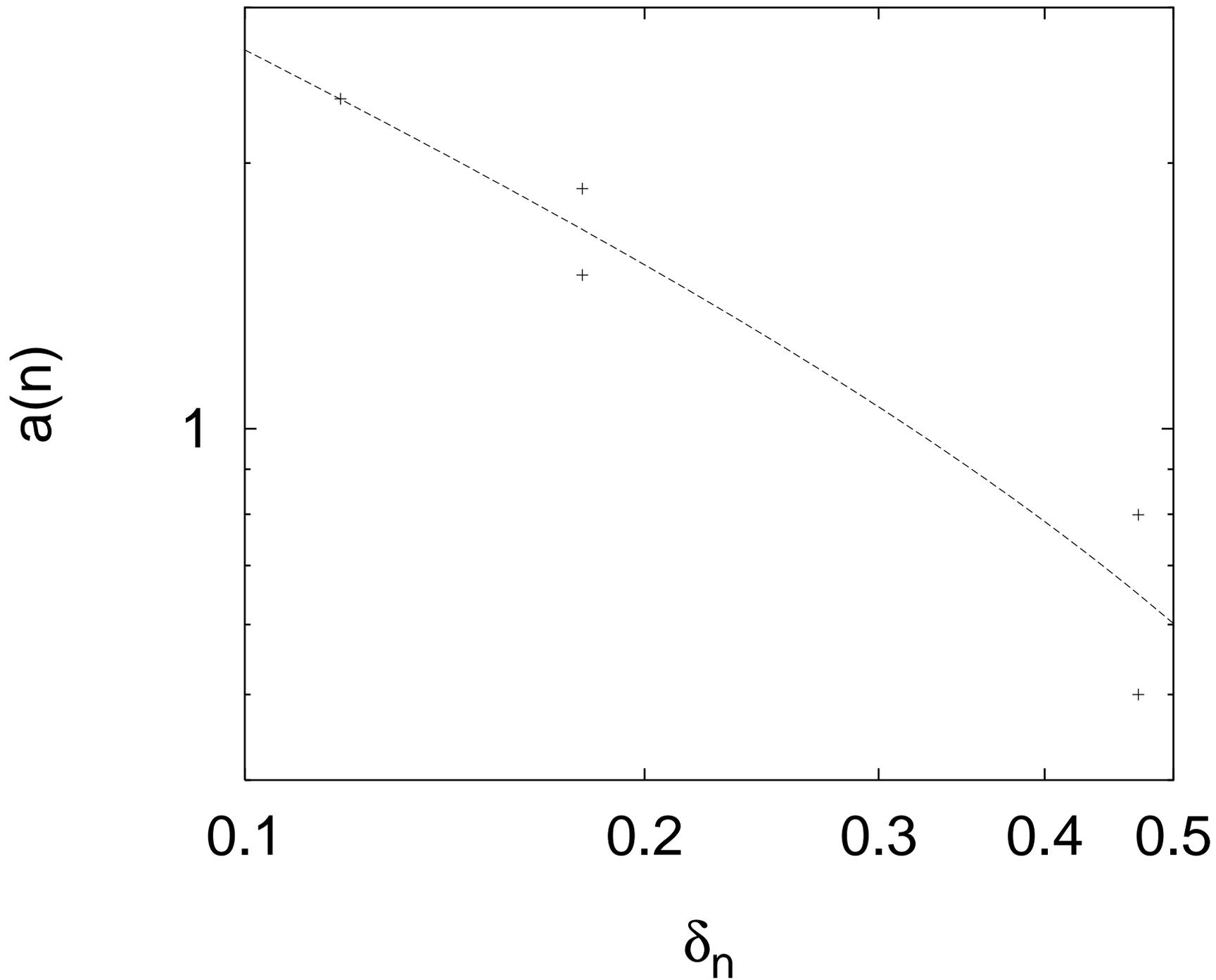
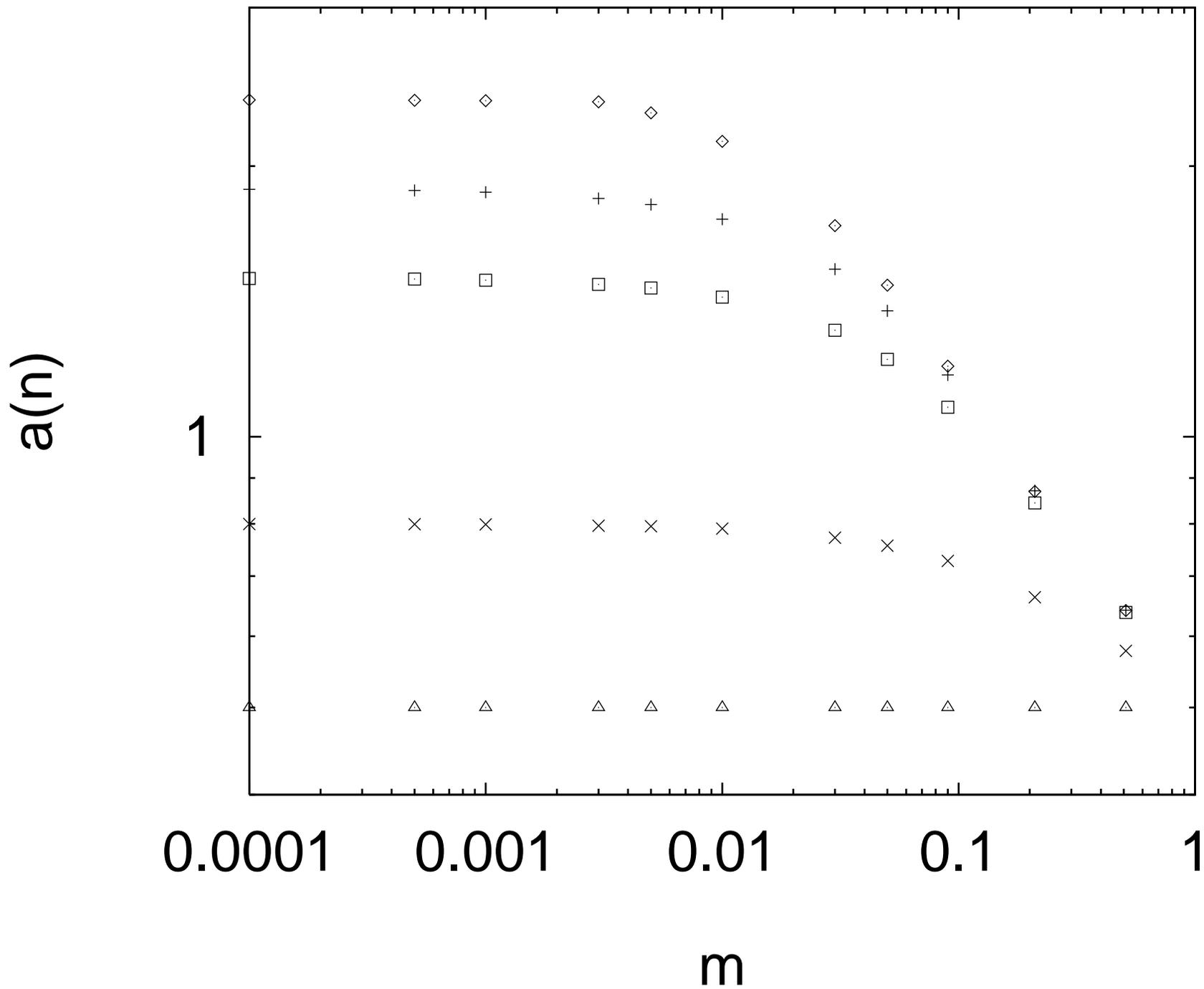
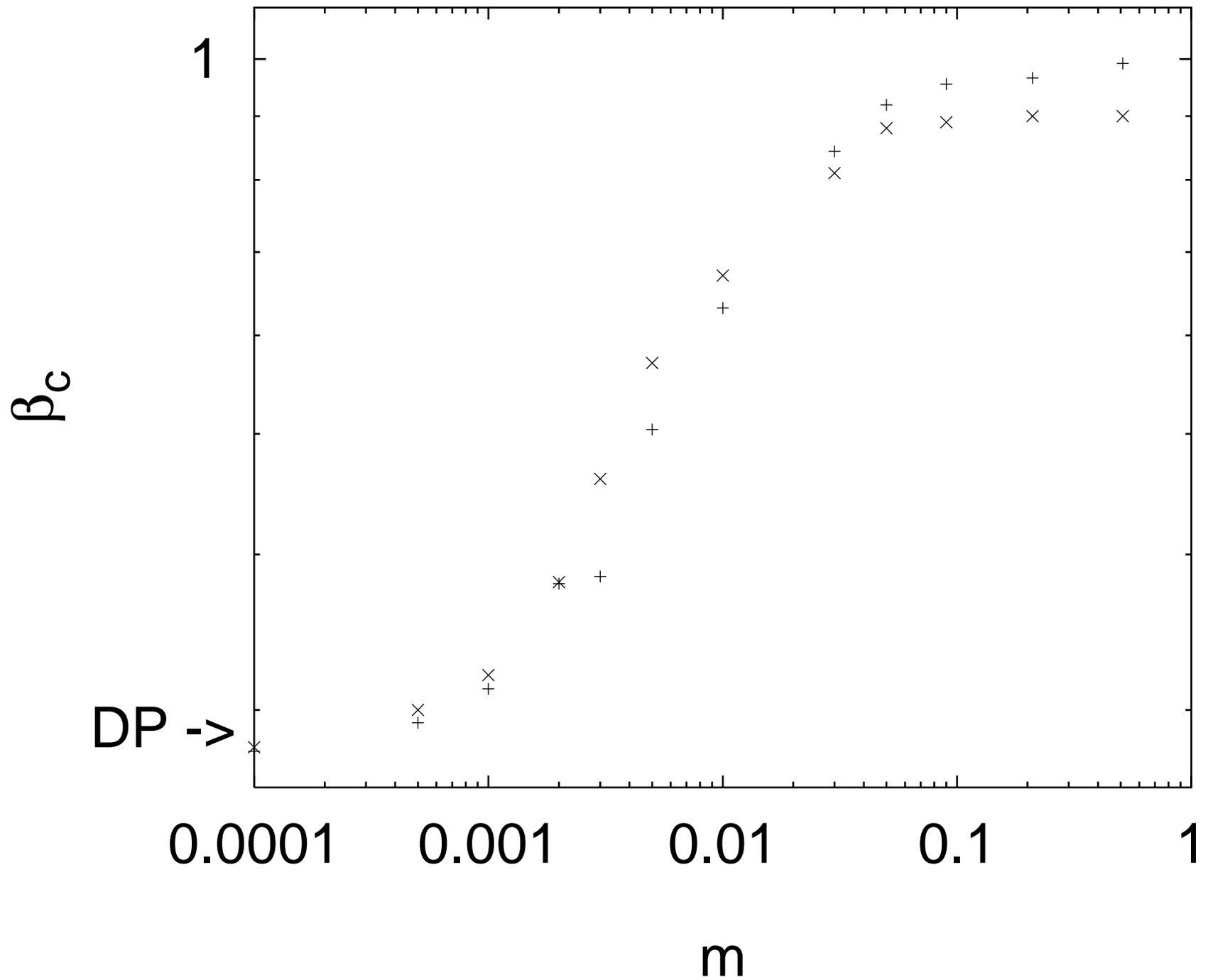
\end{document}